\documentclass[twocolumn]{aastex631}

\usepackage{url}
\usepackage{graphicx}
\usepackage{amsmath,amstext}
\usepackage[T1]{fontenc}

\usepackage{amsmath}
\usepackage{float}
\usepackage{longtable}
\usepackage{rotating}
\usepackage{multirow}
\usepackage{booktabs,tabularx}
\usepackage{CJK}

\graphicspath{{./}{}}

\shorttitle{Double Hot Jupiters through ZLK Migration}
\shortauthors{Liu et al.}

\begin{document}
\begin{CJK*}{UTF8}{gbsn}
\title{The Formation of Double Hot Jupiter Systems through von Zeipel-Lidov-Kozai Migration}


\author[0009-0007-9211-2884]{Yurou Liu (刘雨柔)} 
\affiliation{Department of Astronomy, Yale University, New Haven, CT 06511, USA}

\author[0000-0003-0834-8645]{Tiger Lu (陆均)}
\affiliation{Department of Astronomy, Yale University, New Haven, CT 06511, USA}

\author[0000-0002-7670-670X]{Malena Rice (米乐娜)}
\affiliation{Department of Astronomy, Yale University, New Haven, CT 06511, USA}


\correspondingauthor{Yurou Liu}
\email{yurou.liu@yale.edu}


\begin{abstract}

The von Zeipel-Lidov-Kozai (ZLK) mechanism with tidal friction has been demonstrated as a promising avenue to generate hot Jupiters in stellar binary systems. Previous population studies of hot Jupiter formation have largely examined this mechanism in systems comprised of three bodies: two stars and one planet. However, because stars in a binary system form in similar environments with comparable metallicities, the formation of a single hot Jupiter in such a system may imply that the conditions are more likely met for the companion star, as well. We investigate the ZLK mechanism with tidal friction as a potential mechanism to produce \textit{double} hot Jupiter systems in stellar binaries. Using \textit{N}-body simulations, we characterize the evolution of two cold Jupiters, each orbiting one star in a binary system, undergoing mirrored ZLK migration. We then examine the robustness of this mechanism to asymmetries in stellar masses, planet masses, and planet orbital inclinations relative to the binary plane. We predict that, under the assumptions that (1) most hot Jupiters in binary star systems form through ZLK migration of primordially formed cold Jupiters and (2) if one star in a binary system forms a cold Jupiter, the second does as well, a comprehensive search could identify double hot Jupiters in up to $\sim$9\% of the close- to moderate- separation ($a_*\leq2000$ AU) binary systems that already host a known hot Jupiter. We also argue that a blind search for ZLK-migrated double hot Jupiters should prioritize twin stellar binaries with pericenter approaches of a few hundred AU.

\end{abstract}


\vspace{-10mm}
\keywords{hot Jupiters (753), exoplanet dynamics (490), dynamical evolution (421), exoplanets (498), exoplanet systems (484), exoplanet evolution (491)}

\section{Introduction} 
\label{section:intro}

Hot Jupiters -- giant exoplanets with masses $0.3M_\mathrm{Jup}\leq m_\mathrm{p}\leq13M_\mathrm{Jup}$ and pericenter distances $q\equiv a_\mathrm{p}(1-e_\mathrm{p})<0.1\mathrm{AU}$ for semimajor axis $a_\mathrm{p}$ and orbital eccentricity $e_\mathrm{p}$ -- comprise a relatively rare class of planets that reside around roughly 1\% of FGK stars \citep[e.g.][]{wright2012frequency, howard2012planet, wittenmyer2020cool, beleznay2022exploring}. Previous work has indicated that hot Jupiters may form through three general pathways: in situ formation, disk migration, or high-eccentricity migration \citep[e.g.][]{dawson2018origins}. Though other formation mechanisms are feasible, secular evolution followed by high-eccentricity migration has been demonstrated as a natural and relatively favored avenue for hot Jupiter formation \citep[e.g.][]{wu2003planet, naoz2012formation, vick_2019, rice2022origins, zink2023hot}. This formation avenue requires the presence of another perturbing body such as an additional planet or a stellar companion, and indeed many hot Jupiters have been found in binary star systems \citep{piskorz2015friends, ngo2016friends} or with potential wide planetary companions \citep[e.g.][]{bakos2009hat, neveuvanmalle2016hot,zink2023hot}.

The secular evolution of hierarchical misaligned three-body systems -- for instance, a binary star system with a planet around one of the stars -- can be described through the von Zeipel-Lidov-Kozai \citep[ZLK;][]{von_zeipel_1910, lidov1962evolution, kozai1962secular, naoz2016eccentric} mechanism. Orbital precession of the inner planet driven by the outer perturber results in angular momentum exchange between the two orbits, triggering coupled high-amplitude oscillations in the eccentricity and inclination of the planet's orbit. Another relevant physical effect is tidal dissipation, which affects hot Jupiter progenitor systems in two ways. First, orbital energy is drained from the planet during each pericenter approach, decreasing its semimajor axis and drawing the planet onto a tighter orbit. Second, tidal dissipation provides another source of orbital precession, increasing in relative importance as the planet's orbit shrinks. Eventually, the planet inspirals onto an orbit small enough such that tidal or general relativistic precession \citep[e.g.][]{einstein1916grundlage, sterne1939apsidal, wu2003planet, fabrycky2007shrinking,naoz2013secular, liu2015suppression} dominates over ZLK precession. This suppresses the ZLK eccentricity and inclination oscillations, and tides then work to efficiently circularize the orbit. Taken together, this is the process of \textit{ZLK migration}, and the end result is a planet on a close-in circular orbit. This mechanism has been invoked as a favorable formation scenario for many hot Jupiters and other close-in planets \citep[e.g.][]{wu2003planet,naoz2011hot,beust_2012, petrovich_2015, Lu_2024}.

Previous studies of ZLK migration for hot Jupiter formation have generally focused on the evolution of a single giant planet around one of the two stars in a binary \citep[e.g.][]{naoz2011hot, anderson2016formation}. Given that in binary systems, both stars form in similar environments with comparable metal enrichment \citep[e.g.][]{duchene2013stellar, offner2016turbulent, sadavoy2017embedded}, one may expect that if the conditions for hot Jupiter production are met for one star in a binary system, they may be more likely met for the second star. Indeed, one such \textit{double} hot Jupiter system -- WASP-94 -- has already been confirmed \citep{neveu-vanmalle2014wasp94}.

In this work, we examine the efficiency of double hot Jupiter formation through simultaneous ZLK high-eccentricity migration around each star in a binary system. The ZLK effect in these so-called ``2+2'' systems (where the traditional single outer perturber is replaced by a binary system) has been examined both numerically \citep{pejcha_2013} and analytically \citep[e.g][]{hamers_lai_2017, hamers_2021, oconnor_2021, klein_katz_2024} in previous work. However, the complexity of the four-body system necessitates approximations -- for instance, assuming the central body is much more massive than the binary perturber. Our models fall outside of this regime, so we elect to use \textit{N}-body simulations to capture the relevant dynamics.
The aim of this work is not to study the general four-body dynamics of quadrupole ZLK systems. For our systems, the dynamics are expected to be qualitatively similar to the traditional three-body case due to the mass ratio differential between the host star and its planetary companion. Rather, in this work we present a population-level analysis of this mechanism's efficiency in forming double hot Jupiter systems by approximating the system dynamics as that of classic ZLK migration, and by numerically characterizing perturbation-induced deviations from this approximation.

We first argue that, in a perfectly mirrored system, double hot Jupiters naturally form through double ZLK migration. Given that real systems will not be perfectly mirrored, we then examine the robustness of this mechanism across a range of mutual orbital inclinations and mass asymmetries. Assuming isotropic initial orbital orientations and hot Jupiter formation through ZLK migration, we ultimately find that double hot Jupiter systems may arise through ZLK migration in up to $\sim$9\% of the known close- to moderate- separation ($a_*\leq2000$ AU), \textit{Gaia}-resolved hot-Jupiter-hosting binary systems.

\section{Symmetric System Proof of Concept}
\label{section:archetype}
In this section, we use \textit{N}-body simulations to investigate a fiducial test case of a perfectly mirrored, symmetric system as a proof-of-concept that the dynamical conditions for double ZLK migration are easily satisfied.

\subsection{The Test Particle Approximation}
\begin{figure}
    \centering
    \includegraphics[width=0.48\textwidth]{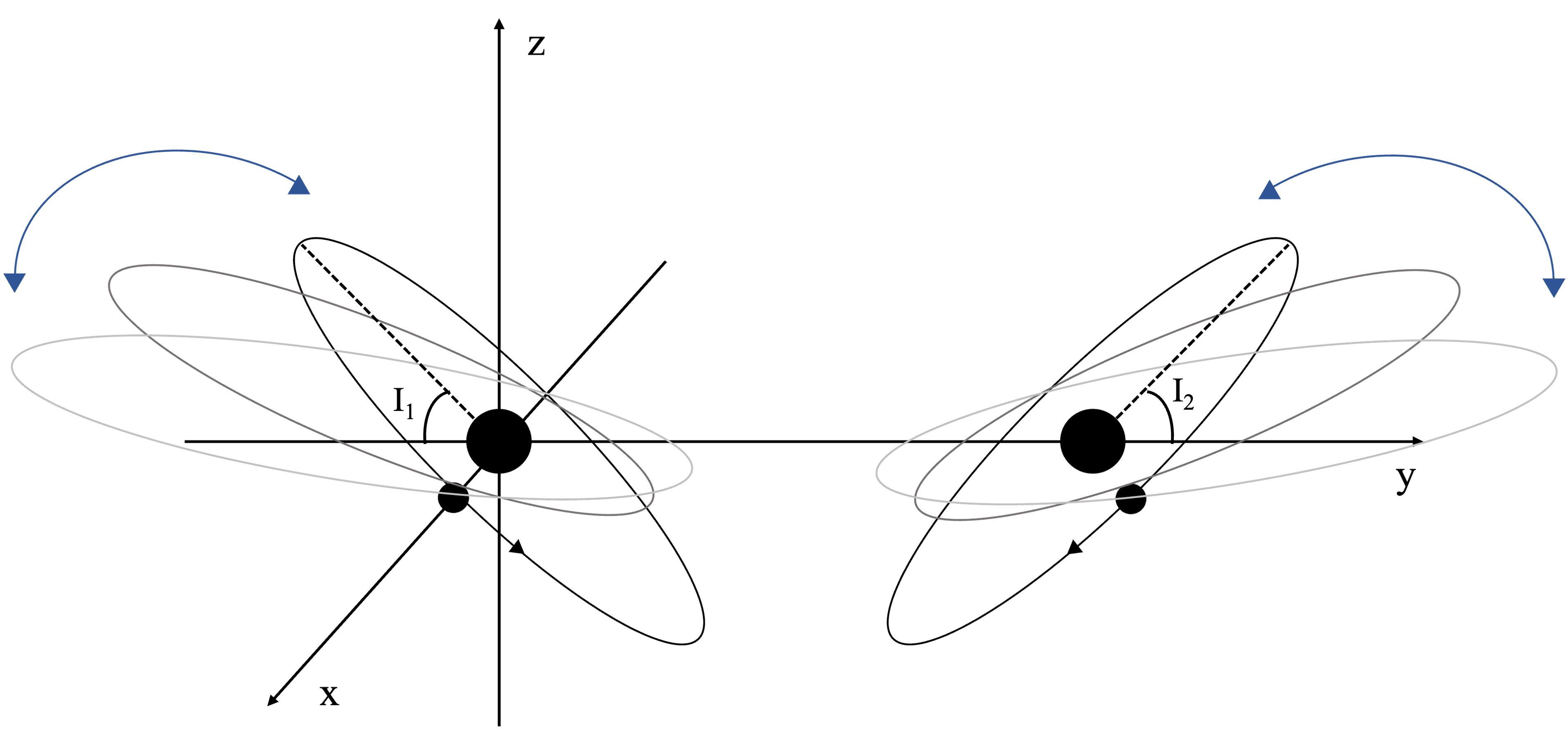}
    \caption{Schematic of the idealized, symmetric binary system with planetary orbital inclinations $I_1 = I_2$. The two planets have the same initial mass, radius, orbital semimajor axis, and orbital eccentricity, and the mutual inclination of the planets with respect to the binary plane adds up to $180^{\circ}$. The stars also have the same mass and radius as each other. The symmetric binary system is the most optimal for double hot Jupiter formation, as demonstrated in Section \ref{section:asymmetries}.}
    \label{fig:schematic}
\end{figure}

We first consider the simplest level of approximation in which the two stars in a binary system are endowed with mass, but the planets around each star are not. The setup of such a symmetric system is shown in Figure \ref{fig:schematic}, where the stellar masses, planetary semimajor axes, and planetary orbital eccentricities are the same, and the mutual inclinations of the planets' orbits with respect to the binary plane add up to $180^{\circ}$. In this case, the two planets have no gravitational influence on each other, and they evolve fully independently. Two cold Jupiters initialized in a mirrored configuration will undergo exactly the same secular orbital evolution: if one planet meets the conditions required to become a hot Jupiter through ZLK migration, then the other does, as well. 

In the limit of perfect symmetry and negligible planet masses, each planet's evolution may be modeled independently. Real systems, even when perfectly symmetric, do not exactly satisfy this approximation, since each planet is endowed with mass. However, as we show in Appendix \ref{appendix:appendix_perturbations}, this approximation is acceptable in the case that the two stars have sufficiently wide-separation close approaches.

\subsection{A Fiducial Mass-Endowed System}
\label{section:symmetric-sim}
Next, we maintain the symmetry of our fiducial system but loosen the assumption that the two planets within the system are massless. We employ \textit{N}-body simulations to directly demonstrate the evolution of mirrored systems in which the two Jupiters are both endowed with mass, qualitatively showing that the two Jupiters' orbits evolve in a near-mirrored manner.

Our fiducial four-body system is initialized with conditions summarized in Table \ref{tab:fiducial}. Orbital parameters were selected to produce a relatively rapidly-evolving (moderate stellar separation, high-eccentricity, and planetary inclinations close to $90^\circ$) system for our proof-of-concept, with mirrored inclinations relative to $I=90^\circ$. Physical properties were drawn from \citet{wu2003planet}, except for the rotation periods of the stars and planets. The planets' rotation periods are analogous to values used in \cite{lu2023tidespin}, while the stars' rotation periods are comparable to the observed values for FGK stars \citep{2014ApJS..211...24M, Colman2024stellarrotation}.

We simulated the evolution of this fiducial system using \texttt{REBOUND} \citep{rein2012rebound} with the high-order adaptive-timestep \texttt{IAS15} integrator \citep{rein2015ias15, pham2024new}, which allows us to resolve close pericenter approaches with the star during epochs of high eccentricity. General relativity and tides were incorporated using the \texttt{REBOUNDx} package \citep{tamayo2019reboundx}. The effects of general relativity were incorporated using the \texttt{gr\_full} prescription \citep{Newhall1983grfull}, which accounts for first-order post-Newtonian effects from all bodies in the system. We also incorporate equilibrium tides \citep{eggleton1998tidespin} with the \texttt{tides\_spin} implementation \citep{lu2023tidespin}, which allows for self-consistent spin and dynamical evolution under the influence of tidal friction. We set the value of the tidal quality factor to $Q_{\rm HJ}=10^3$ to speed up our simulations, making the planet a factor of $300$ more dissipative than the typical literature value $Q_{\rm{HJ}} = 3 \times 10^5$\citep{wu2003planet}\footnote{The equilibrium tide model implemented by \cite{lu2023tidespin} uses the constant time lag $\tau$ as the tidal parameter. We convert between $\tau$ and $Q$ with the commonly used conversion $\tau = 1/(2 n Q)$, where $n$ is the mean motion. In our simulations, we adopt $n$ appropriate for a 10 day orbit.}. This is a common practice in numerical simulations \citep[e.g][]{bolmont_2015, Becker20,faridani2024likely, Lu_2024} as it does not affect the period or amplitude of the ZLK cycles, but instead serves only to speed up our simulation runtimes by increasing the energy dissipated during each pericenter approach. We rescale linearly to recover and report realistic timescales in our figures.


\begin{table}
\centering
\begin{tabular}{lll}
\toprule
Parameter  & Definition & Value\\\midrule
$M_{*,i},m_i$ & Stellar/planetary masses & $1 M_\odot, 1 M_{\mathrm{Jup}}$ \\
$R_{*,i},r_i$ & Stellar/planetary radii &$1 R_\odot, 1 R_{\mathrm{Jup}}$\\
$a_*,a_i$ & Semimajor axes & $200, 5$ AU\\
$e_*,e_i$ & Eccentricities & $0.7, 0$ \\
$I_*,I_1,I_2$ & Inclinations & $0^\circ, 83^\circ, 97^\circ$\\
$\omega_*, \omega_i$ & Arguments of periapsis & $90^\circ, 0^\circ$\\
$k_{2,*},k_{2,\mathrm{Jup}}$ & Love numbers & $0.028, 0.51$ \\
$C_*,C_{\mathrm{Jup}}$ & Gyroradii &  $0.08, 0.25$\\
$\Omega_*,\Omega_{\mathrm{Jup}}$ & Rotation periods &  $12$ days, $1$ day\\
\bottomrule
\end{tabular}
\caption{Fiducial values for our \textit{N}-body simulations. All quantities subscripted with $*$ refer to values associated with the stellar hosts; all others refer to the planets. The two stars are identical, as are the two planets. The Love number \citep{Love_yielding} and gyroradius values were adopted from \citet{wu2003planet}. While all parameters are reported for completeness, we note that the most impactful parameters in this study are the masses, planetary radii, semimajor axes, and mutual inclinations.}
\label{tab:fiducial}
\end{table}

\begin{figure}
    \centering
    \includegraphics[width=0.48\textwidth]{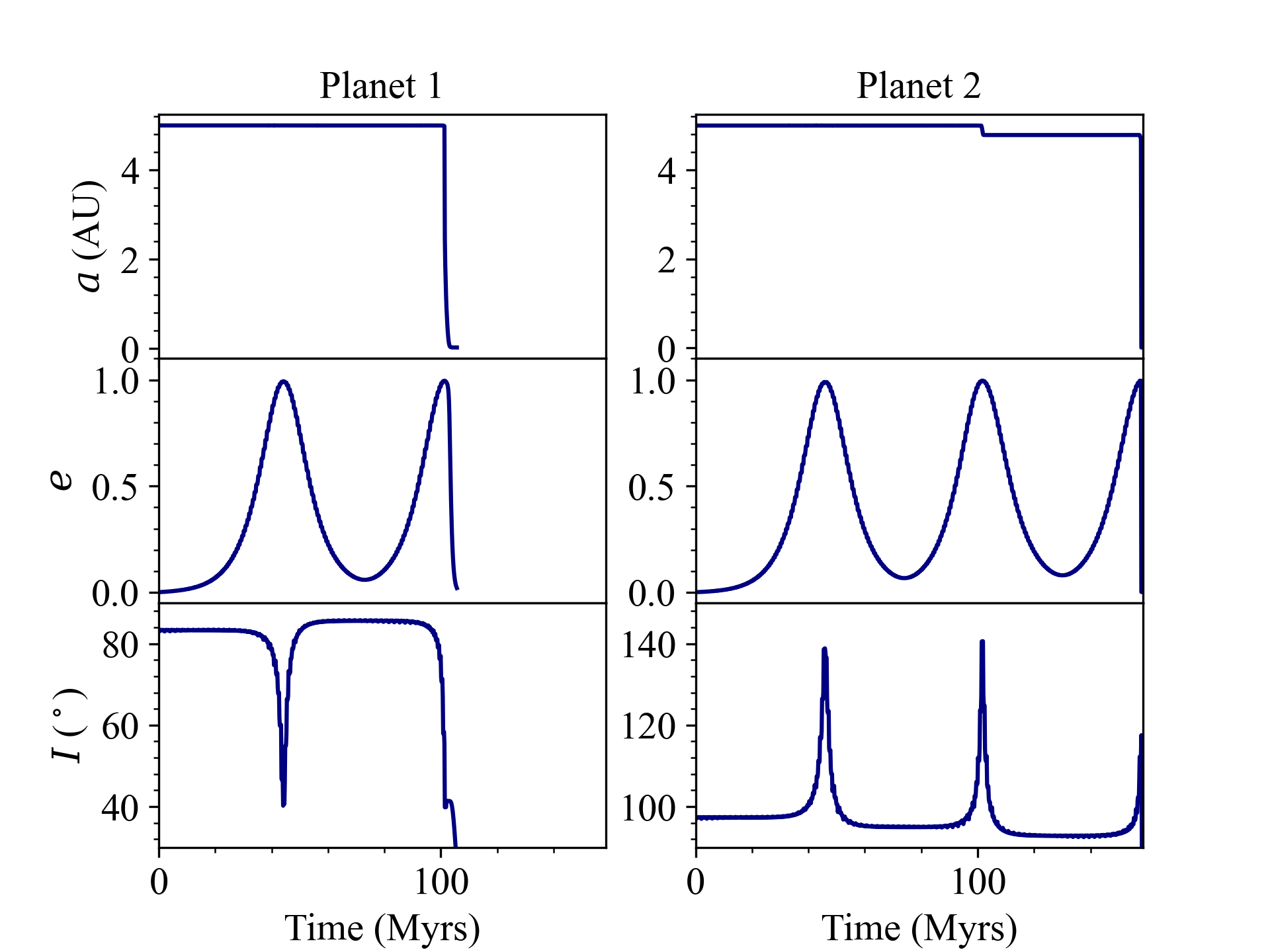}
    \caption{Semimajor axis, eccentricity, and inclination evolution of the two planetary orbits undergoing ZLK oscillations. The left and right panels show the evolution of the first and second planet, respectively. The planet 1 evolution is cut off upon manually colliding the planet with its host star after it has become a hot Jupiter. The continued inclination evolution of planet 1 after ZLK cycles conclude is due to precession of its orbit normal about the invariant plane of the system.}
    \label{fig:symmetric-sim}
\end{figure}
We consider the planet to be a hot Jupiter once it has reached pericenter distance $q<0.1$ AU.\footnote{This $q$ constraint, following \citet{rice2022origins}, was selected to ensure that the hot Jupiters' pericenter distances are sufficiently close to the host star for tidal circularization to have an important effect during the system lifetime.} After one hot Jupiter has formed, the adaptive-timestep $\texttt{IAS15}$ integrator is forced to use extremely small timesteps to resolve the new hot Jupiter's short-period orbit. Such small timesteps prohibitively increase the integration time required to further evolve the second planet that is still undergoing ZLK oscillations. To evade this bottleneck, after the first hot Jupiter forms (reaching $q<0.1 \mathrm{AU}$) and circularizes, it is manually collided with its host star through an inelastic collision, conserving mass and angular momentum. The second planet's evolution then continues to be integrated with this combined object -- an approximation to the first hot Jupiter system -- as the perturber. This approximation maintains the total mass of the perturbing body with respect to the remaining planet.

The results of the simulation are shown in Figure \ref{fig:symmetric-sim}. In our fiducial system, both planets evolved to become hot Jupiters with close-in circular orbits within ${\sim}2\times10^8$ years. We show in Appendix\,\ref{appendix:appendix_perturbations} that, when the three bodies have sufficiently wide close approaches, planet-planet perturbations become less meaningful in influencing the system's evolution.

We note that, in many theoretical studies of hot Jupiter formation via ZLK migration \citep[e.g.][]{naoz2011hot, petrovich_2015}, a subset of potential hot Jupiters are perturbed onto eccentric orbits with pericenters too close to their host stars and are thus tidally disrupted. The tidal disruption limit corresponds roughly to the Roche limit, which is given as \citep{matsumura_2010_tidal, naoz2012formation}

\begin{equation}
    a_\mathrm{Roche} = \frac{1}{0.6} r_\mathrm{p}\Big(\frac{m_\mathrm{p}}{M_* + m_\mathrm{p}}\Big)^{-1/3} \sim 0.008 \mathrm{AU}.
    \label{eq:r_roche}
\end{equation}
We check whether the planets cross the Roche limit at each timestep and flag cases in which a planet is tidally disrupted. In this simulation, both planets approached but did not cross the Roche limit, so no tidal disruption events were recorded.

\section{The Impact of Asymmetries}
\label{section:asymmetries}
Realistic planetary systems include asymmetries in their inclinations and/or their masses. In this section, we relax the assumption of a perfectly mirrored system and quantify how these asymmetries affect our results. We find that inclination asymmetries have the most pronounced impact, as the initial inclination of the system informs the initial maximum eccentricity of the ZLK cycles and hence largely dictates the hot Jupiter formation rate. Highly unequal stellar masses may shift hot Jupiter formation timescales by up to a factor of a few, but they otherwise do not qualitatively affect the systems' evolution.

\subsection{Inclination Asymmetries}
The rate of ZLK migration is highly sensitive to the initial mutual inclination between the stellar binary orbit plane and the planetary orbit plane \citep{naoz2012formation}. If the initial mutual inclination is very low, ZLK oscillations are never initiated. If ZLK oscillations do begin, the initial mutual inclination informs the initial maximum amplitude of eccentricity oscillations and hence the timescale of evolution toward hot Jupiter formation. Therefore, we first investigate how asymmetric initial inclinations affect the success rate of hot Jupiter formation in our fiducial setup. We note that, observationally, exoplanet-hosting binary systems show a suggestive trend toward low mutual inclinations \citep{christian2022possible, dupuy2022orbital, rice2024orbital} that may reflect dissipative precession at the protoplanetary disk stage \citep{gerbig2024aligning}, damping inclinations and preventing the initialization of ZLK oscillations. This trend is not clearly present across the full set of hot-Jupiter-hosting binary systems \citep{christian2024wide}, so we assume isotropy within this work.

We use the same initial conditions described in Section \ref{section:symmetric-sim}, but we consider an isotropic $\cos{I_1}$ distribution of 59 values from -1 to 1 (inclusive). Once the pericenter distance of both planets satisfies $q<0.1\mathrm{AU}$, the configuration is deemed ``successful'' in producing a double hot Jupiter system, and the simulation is halted. We observe that when $|\cos{I_1}| > 0.689$, the planet does not evolve into a hot Jupiter.\footnote{Note that this threshold is specific to the configuration examined here. The exact mutual inclination threshold varies with the mass, semi-major axis, and eccentricity of the binary companion, as well as physical properties of the planet such as tidal quality factor, Love number and radius  \citep[e.g.][]{naoz2012formation, Storch2014tides, petrovich_2015, vick_2019}.} To more clearly resolve the threshold for the formation of hot Jupiters in this configuration, we perform another suite of simulations with $I_1$ uniformly varied across $1500$ values from $-0.689 < \cos{I_1} < 0.689$. Using these simulations, we derive the success rate of double hot Jupiter production with our setup.


\begin{figure}
    \centering
    \includegraphics[width=0.48\textwidth]{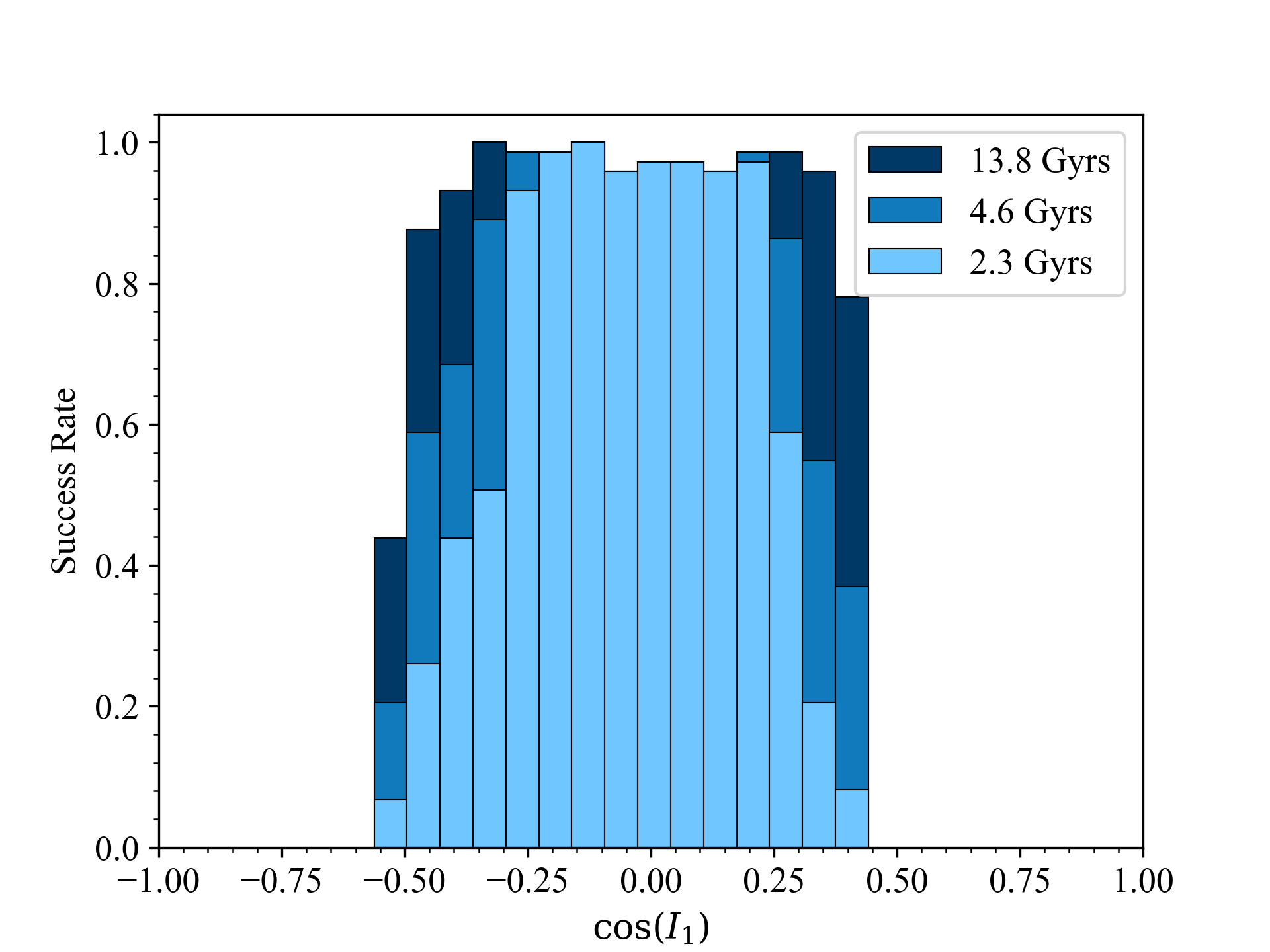}
    \caption{Success rates of double hot Jupiter production as a function of $\cos I_1$. The rates are calculated from an isotropic sample of numerical simulations initiated uniformly in $\cos I_1$. The provided times are in ``real time,'' translated from simulation time. The binary semimajor axis, binary eccentricity, and initial planet eccentricities are set to $a_*=200$ AU, $e_*=0.7$, and $e_{1,2}=0$ for this set of simulations.}
    \label{fig:inclination-grid}
\end{figure}
In this regime, 38 runs were halted because at least one of the two planets crossed the tidal disruption radius (Equation \ref{eq:r_roche}). In the remainder of the simulations, one of the planets always evolves into a hot Jupiter during the lifetime of the system due to a very high initial mutual inclination. For an isotropic set of inclinations for the second planet, the overall success rate of producing double hot Jupiters in the lifetime of the universe ($13.8$ Gyr) is $\sim$46\%. However, most hot-Jupiter-hosting systems are not as old as the universe \citep{chen2023young}, so we also calculate the overall rates of double hot Jupiter production within a third and a sixth of this time limit (4.6 Gyr and 2.3 Gyr, respectively). These success rates are $\sim$40\% and 33\%, respectively. The double hot Jupiter formation rates corresponding to each time limit are shown in Figure \ref{fig:inclination-grid} as a function of $\cos{I_1}$. 



\subsection{Mass Asymmetries}

The masses of the stars and/or planets can also be unequal in a binary system. To investigate the effect of mass asymmetries, we again run a suite of \textit{N}-body simulations, varying the mass of either one star or one planet to separately parse each effect.

\begin{figure}
    \centering
    \includegraphics[width=0.48\textwidth]{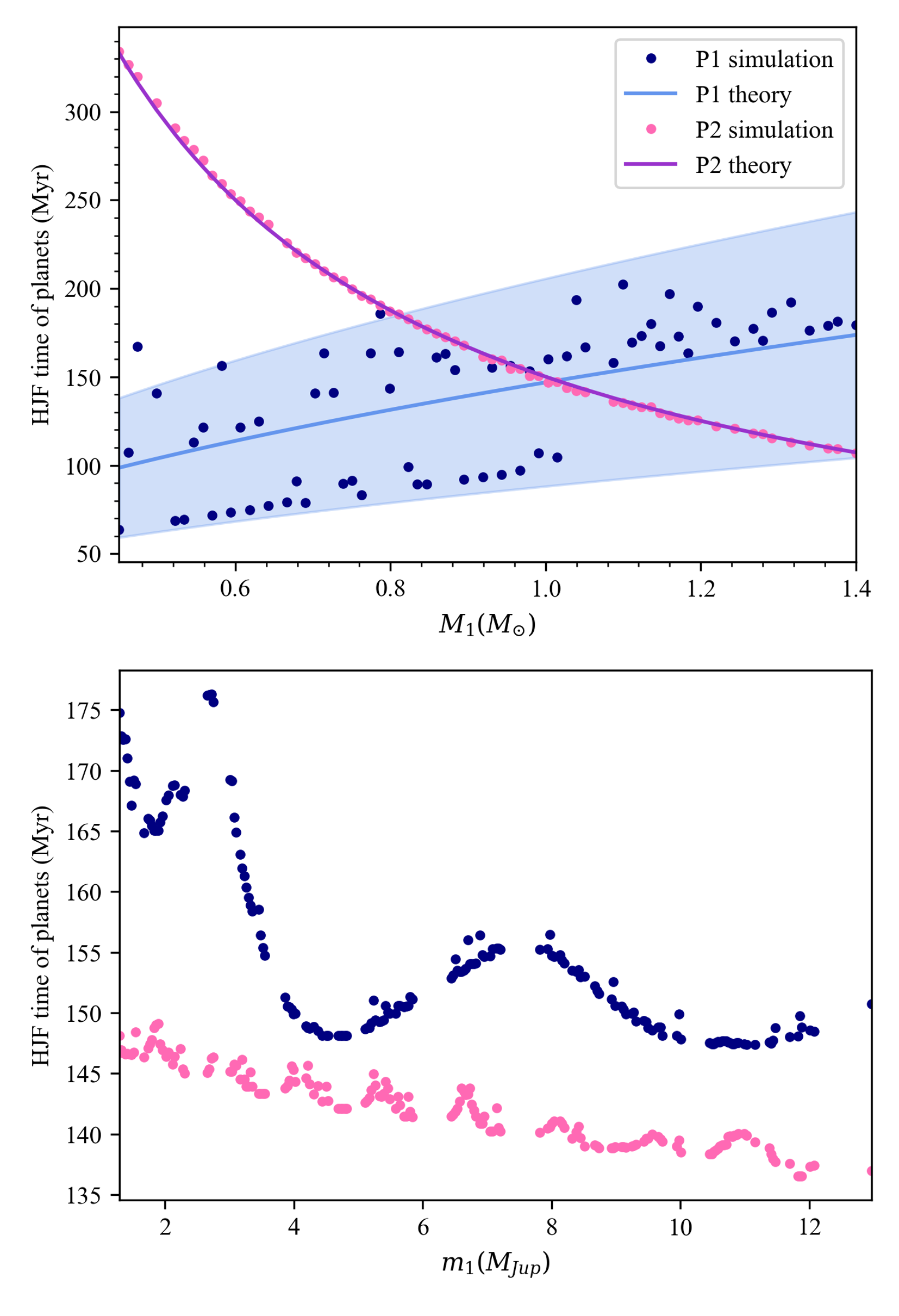}
    \caption{The hot Jupiter formation (HJF) time, in real time, of 4-body double hot Jupiter simulations with asymmetric masses. System parameters are adopted from Table \ref{tab:fiducial} except for the masses, which we vary. \textit{Top:} Change in hot Jupiter formation timescale with changing stellar mass. Theoretical predictions based on the ZLK timescale are overplotted. The shaded region spans a $\pm1$ ZLK cycle error bar as defined in Equation \ref{eq:t_zlk}. \textit{Bottom:} Same as the top panel, but with changing planetary mass. 10 systems with anomalously short or long formation times, due to chaotic behavior, have been excluded to more clearly highlight the overall trend.}
    \label{fig:mass-asym}
\end{figure}

\subsubsection{Stellar Mass Asymmetry}
\label{subsubsection:star-asymmetries}
To quantify the influence of stellar mass asymmetry on hot Jupiter formation, we ran a suite of simulations with the fiducial parameters in Table \ref{tab:fiducial}, but with 80 uniformly sampled values of $M_1$ between 0.45 ${M_{\odot}}$ and 1.4 ${M_{\odot}}$ (the mass range characteristic of FGK stars; see e.g. \citet{habets1981stellarmass}). The stellar radius is correspondingly scaled according to the standard mass-radius relation for main sequence stars $R_* \propto M_*^{0.8}$ \citep[e.g.][]{demircan1991stellar}. We use the same criterion $q<0.1$ AU to determine whether a planet becomes a hot Jupiter. 

In 67 out of 80 runs, both planets evolve into hot Jupiters. In the other 13 runs, at least one of the two planets crosses the Roche limit (Equation \ref{eq:r_roche}). Figure \ref{fig:mass-asym} shows the hot Jupiter formation time for each simulation in dark blue (planet 1) and pink (planet 2) points. The hot Jupiter formation time varies substantially with stellar mass, largely due to the ZLK timescale's dependence on stellar mass. 

There is no well-defined fully analytical ZLK timescale that appropriately describes systems with highly eccentric outer perturbers. However, significant insight may still be drawn through comparison with the ZLK timescale in the ``test particle quadrupole'' limit, which describes an inner test particle with an outer perturber on a nearly circular orbit \citep{antognini2015timescales, naoz2016eccentric}:


\begin{equation}
    t_{\rm ZLK, i} \propto \frac{M_{\mathrm{tot}}}{M_{*,i}+m_{i}}\frac{P_{*}^2}{P_{i}}(1-e_*^2)^{3/2},
    \label{eq:t_zlk}
\end{equation}
where $i \in [1,2]$. $P_1$ and $P_2$ are the orbital periods of the two planets, $P_*$ is the orbital period of the stellar binary, and $M_{\mathrm{tot}} = m_1+m_2+M_{*,1}+M_{*,2}$ is the total mass of the system. To demonstrate the first-order relationship between the hot Jupiter formation time and the ZLK timescale, we scale Equation \ref{eq:t_zlk} by the ratio of the average simulation timescale to the average analytic value. This captures the relative scaling across different stellar masses while also accounting for order-unity differences in expected prefactors between the well-defined timescale for a circular perturber given in Equation \ref{eq:t_zlk} and the eccentric case considered in our simulations. 

The scaled version of Equation \ref{eq:t_zlk} is overplotted in Figure \ref{fig:mass-asym}. While the trend of hot Jupiter formation time is reasonably well described by the ZLK timescale, significant scatter remains. We interpret that this is caused by significant octupole-order effects in our simulations, caused by a high $e_*$. 

The evolution of the system is chaotic \citep{naoz2013secular, li14chaos}, such that small changes in the planet's maximum eccentricity profoundly impact the rate of tidal dissipation in the planet's orbit, which scales as $e_{1,2}^2$ \citep{leconte2010tidal, millholland2019obliquity}. Hence, the number of ZLK cycles required to generate a hot Jupiter varies unpredictably. In our setup, planet 1 is more clearly affected by this chaos than planet 2, since we vary only $M_{*,1}$. More chaotic behavior would be expected for planet 2 in the case that $M_{*,2}$ is varied as well.


We now discuss the general trends illuminated by Figure \ref{fig:mass-asym}. As $M_{*,1}$ increases, the first system's hot Jupiter formation time increases while the formation time of the companion's hot Jupiter decreases. The time taken for both planets to become hot Jupiters is minimized when $M_{*,1} \simeq M_{*,2} \simeq M_{\odot}$, showing that double hot Jupiters form most efficiently in systems where the stellar masses are equal. As $M_{*,1}$ decreases below $M_{*,2}$, the hot Jupiter formation time around star 2 increases significantly and the total time taken to produce a hot Jupiter can be as much as $2.2\times$ longer than in the symmetric system. In some cases, therefore, the time required for asymmetric systems to produce double hot Jupiters may be longer than the system's lifetime. 

Figure \ref{fig:mass-asym} also shows that the hot Jupiter formation time is longer around the primary (more massive star) than around the secondary. As such, if a hot Jupiter is detected around the primary star and if one could eventually form around the secondary star, the relative timescale favors it having already formed, assuming that both hot Jupiters are produced through binary-induced ZLK migration. If a hot Jupiter is found around the secondary star, then even if a hot Jupiter can form around the primary given enough time, it might not have done so yet. The formation time increases as the host's mass (primary mass) increases and the perturber's mass (secondary mass) decreases, as seen from inspection of Equation \ref{eq:t_zlk}. In this case, the probability of seeing the second hot Jupiter is highest when the masses of the stars are closest.


\subsubsection{Planetary Mass Asymmetry}
To investigate the influence of planetary mass asymmetries on double hot Jupiter formation, we ran an analogous series of simulations with both stellar masses held constant while varying the mass of planet 1. The mass of planet 1 was uniformly sampled across 400 values within the range $0.3M_\mathrm{Jup}\leq m_{\rm p} \leq 13 M_\mathrm{Jup}$, a fiducial mass range for hot Jupiters \citep[e.g.][]{winn2010hot, dawson2018origins, yee2021how}. The planet's radius was kept constant, as the radius of Jovian planets is relatively insensitive to mass \citep[e.g.][]{weiss2013mass, hatzes2015radius, chen2017radius, muller2024radius}. The same criteria were used to determine whether hot Jupiter formation was successful.

Of the 400 runs, 178 systems produced double hot Jupiters, but the hot Jupiter formation time varied with changing planet mass. In the other 222 runs, as in the 13 unsuccessful runs in the stellar asymmetry simulations (Section \ref{subsubsection:star-asymmetries}), at least one of the two planets was tidally disrupted. We note that to save computing time, our setup of $a_* = 200 \mathrm{AU}$ and $e_* = 0.7$ leads to a large number of planets crossing the Roche limit. This outcome is not representative of more realistic systems with a wider separation. Simulation results are shown in Figure \ref{fig:mass-asym}. 


We found that the simulated hot Jupiter formation time does not follow the scaling of Equation \ref{eq:t_zlk} as had been observed in Section \ref{subsubsection:star-asymmetries} for varying stellar mass. This is most likely due to the significant planet-planet interactions that are not incorporated into Equation \ref{eq:t_zlk} (see Appendix \ref{appendix:appendix_perturbations} for a more detailed description) and the influence of octuple-order effects that dominate the hot Jupiter formation time difference between different runs. As shown in Figure \ref{fig:mass-asym}, changing planet mass alters hot Jupiter formation times on a scale significantly smaller than the variations caused by changing stellar mass, and smaller than the typical age uncertainties of hot Jupiter host stars \citep[e.g.][]{chen2023young}. 

10 outlier points were removed from Figure \ref{fig:mass-asym}, corresponding to systems that have either undergone oscillations that are not well-defined by the classical ZLK timescale or that have undergone one more/fewer cycle than the norm. Since these outliers reflect the chaotic nature of the system rather than systematic behavior, we exclude them to provide a clearer view of the underlying trend. The impact of these effects is overestimated in our simulations, which run for only a few ZLK cycles with artificially dissipative planets to minimize computation costs. In reality, hot Jupiters may undergo hundreds of cycles before reaching their present-day orbits; at such scales, chaotic effects are less relevant.


\section{The Production Rate of Double Hot Jupiters in Twin Binary Systems}
\label{sec:gaia}

Having shown that two hot Jupiters can form in conjunction from simultaneous ZLK oscillations in the same binary system, we aim to deduce the expected occurrence rate of double hot Jupiter systems among systems with one known hot Jupiter. In other words, if one hot Jupiter in a twin (equal-mass) binary star system is produced through the ZLK mechanism, what is the likelihood that a hot Jupiter will be found around the secondary star as well?

This likelihood can be formally quantified through Bayes' theorem as

\begin{equation}
    P(HJ_{1\,\rm{and}\,2} | HJ_1 ) = \frac{P (HJ_1 | HJ_{1\,\rm{and}\,2}) P(HJ_{1\,\rm{and}\,2})}{P(HJ_1)},
\end{equation}
which reduces to 

\begin{equation}
 P(HJ_{1\,\rm{and}\,2} | HJ_1 ) = P(HJ_2)
\end{equation}
when considering that $P (HJ_1 | HJ_{1\,\rm{and}\,2}) = 1$ by definition, and $P(HJ_{1\,\rm{and}\,2})=P(HJ_1) P(HJ_2)$ for independent probabilities of each planet becoming a hot Jupiter. For twin binaries, all composition or stellar environment-based requirements for hot Jupiter formation should be met for the second star if they are met for the first. Therefore, we focus on the likelihood that the second planet will become a hot Jupiter as a function of only its relative geometry. We note the caveat that it is as-yet unclear the extent to which the outcomes of planet formation are correlated across the two components of twin binary star systems \citep[e.g.][]{hand2025case}.

We begin by considering binary systems with one identified hot Jupiter around one star, and we assume that a cold Jupiter formed around the other star. We then conduct a series of $N$-body simulations with \texttt{REBOUND} to determine whether two hot Jupiters could form in these systems.

To draw from a realistic set of binary separations representative of known hot Jupiter systems, we considered the set of known hot Jupiters in moderate- to wide-separation binary system resolved by the \textit{Gaia} mission \citep{gaia2016}. We downloaded the NASA Exoplanet Archive PSCompPars table on June 20th, 2024 \citep{PSCompPars} and filtered to systems that included a planet satisfying the criteria $0.3M_\mathrm{Jup}\leq m_{\rm p} \leq 13 M_\mathrm{Jup}$ and pericenter distance $q<0.1\mathrm{AU}$. The selected hot Jupiter hosts were then cross-matched with the \textit{Gaia} DR3 catalogue \citep{gaia2023dr3} following the methods of \citet{el2021million} to identify bound stellar companions. In total, we identified 94 systems -- 83 of which were also in the \citet{el2021million} catalogue that used \textit{Gaia} eDR3 \citep{gaia2021edr3}, and that excluded triple-star systems.

\begin{figure*}
    \centering
    \includegraphics[width=\textwidth]{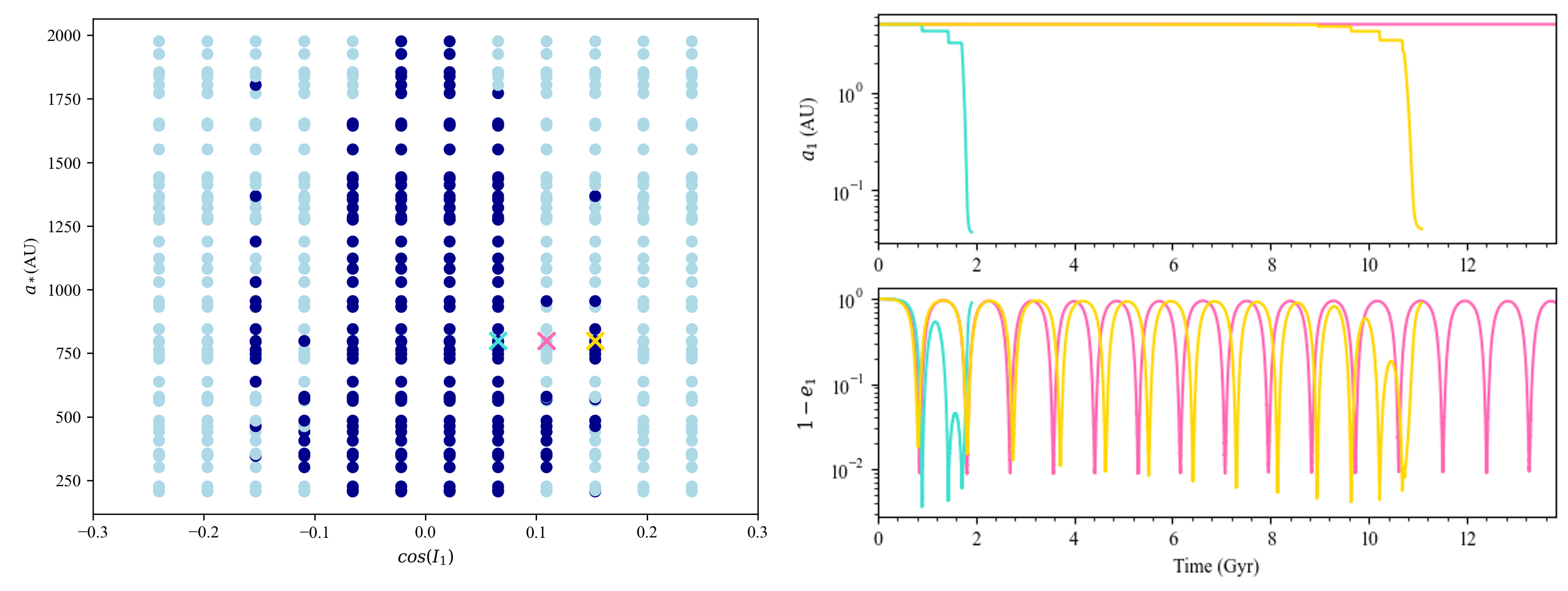}
    \caption{\textit{Left:} Hot Jupiter status after simulating for the age of the universe as a function of stellar binary semimajor axis (set such that $q=200$ AU) and initial $\cos{I_1}$. The dark blue points are hot Jupiters (with $a_1(1-e_1) < 0.1$ AU), whereas the light blue simulations did not successfully evolve into hot Jupiters within the time limit. \textit{Right:} the time evolution of semimajor axis and eccentricity for three specific simulations (marked with the corresponding colored X's in the left subplot). We see that successful formation of hot Jupiters relies on either a high initial maximum eccentricity driven by high initial mutual inclinations (turquoise curve), or octupole-order ZLK effects increasing the maximum eccentricity over time (yellow curve).}
    \label{fig:scatterHJ}
\end{figure*}

Because the ZLK timescale is long at wide binary separations, such that stellar ZLK is unlikely to be the hot Jupiter formation pathway in such systems (see the narrow set of inclinations required for hot Jupiter formation at the top of Figure \ref{fig:scatterHJ}, left panel), we considered only relatively close binaries with sky-projected separation $s\leq2000$ AU. This left 48 systems, of which 46 were also included in the \citet{el2021million} catalogue. 

The ZLK timescale varies with the period and eccentricity of the stellar binary as $t_{\mathrm{ZLK, i}}\propto(P_*^2/P_i)(1-e_*^2)^{3/2}$ (see Equation \ref{eq:t_zlk}) such that, particularly at larger semimajor axes, only systems with high binary eccentricities can form hot Jupiters through ZLK migration within the age of the universe. While individual wide binaries' eccentricities are not well-constrained at scale, statistical studies have demonstrated that average orbital eccentricity increases with binary separation \citep{tokovinin2015eccentricity}. We set the eccentricities of our binaries such that $a_*(1-e_*)=200 \mathrm{AU}$, corresponding to a maximum $e_*\sim0.9$. Because binary eccentricities are not well-characterized for our sample, we adopted $a_*=s$, noting that this will generally be an overestimate of $a_*$ for $e_*\neq0$, where stars will spend most time at apocenter.

We kept the initial inclination $I_2$ fixed at $93^\circ$. The values of $I_1$ were chosen along a grid as shown in Figure \ref{fig:scatterHJ}. We used the same $Q$ scaling described in Section \ref{section:symmetric-sim} to speed up our simulations by a factor of 100 for systems with $a_*$ smaller than $1000\, \mathrm{AU}$ and a factor of 300 for systems with $a_*$ larger than $1000 \,\mathrm{AU}$. The time limit of $N$-body simulations was set to the equivalent of the age of the universe after correcting for the selected factor. The rest of the system parameters, including masses, were set to those shown in Table \ref{tab:fiducial}. To reduce computation time, we used the Gragg-Bulirsch-Stoer (\texttt{BS}) integrator \citep{lu2023tidespin} in \texttt{REBOUND} for this set of simulations. Both tolerances were set to $10^{-10}$. We ran a fiducial system without tidal effects for $4.6\times10^7$ years and found that the energy errors were of order $10^{-6}$.

As shown in Figure \ref{fig:scatterHJ}, most simulations produced double hot Jupiters in the range $-0.153<\cos{I_1}<0.153$. This region is 15.3\% of the parameter space. Within this region, 59.9\% of the systems formed two hot Jupiters, placing an upper limit of $\sim$9\% for the projected occurrence of double hot Jupiters among binary systems with $a_*\leq2000$ AU. This value assumes a relatively small pericenter distance -- in practice, a few hundred AU -- such that ZLK oscillations are efficient.

An interesting feature that merits investigation is the appearance of thin bands in $\cos I_1 - a_*$ space capable of forming hot Jupiters slightly separated in inclination space from the main body of HJ-formation parameter space. Closer inspection of the time-evolution of our simulations (right panels of Figure \ref{fig:scatterHJ}) provides some insight. Extremely high maximum eccentricity is required to efficiently generate a hot Jupiter within the age of the universe. Such high eccentricities can be generated in one of two ways. The first is to simply initialize a system with an extremely high initial inclination, which is directly linked to the maximum amplitude of the ZLK eccentricity oscillations in the quadrupole approximation via

\begin{equation}
    e_\mathrm{p,max} = \sqrt{1 - \frac{5}{3}\cos^2 I_\mathrm{p,initial}}.
\end{equation}
This condition is satisfied by the simulations near $\cos I_1=0$ in the left panel of Figure \ref{fig:scatterHJ}. In the absence of an initial system configuration that is immediately capable of generating high maximum eccentricity, the other way to efficiently create a hot Jupiter is to consider octupole-level effects. At the octupole level of approximation, the maximum eccentricity of a ZLK cycle does not stay fixed -- rather, it slowly grows over time \citep{naoz2013resonant, li14chaos}. Simulations that successfully create hot Jupiters in the bands of lower initial inclination (e.g. the yellow curve in Figure \ref{fig:scatterHJ}) hence correspond to regions of parameter space in which octupole-order effects are relevant. 

The impact of octupole-order effects scales as $(a_1/a_*) [e_*/(1-e_*^2)]$, which strictly increases with stellar semimajor axis in Figure \ref{fig:scatterHJ}, since we have fixed stellar pericenter distance. These bands of lower-inclination hot Jupiter formation are likely truncated to smaller stellar semimajor axes due to interplay with the ``secular descent'' timescale \citep{weldon2024analytic}, which governs how quickly the maximum eccentricity grows due to octupole order effects. This timescale itself scales with the ZLK timescale, which increases with semimajor axis -- see Equation \eqref{eq:t_zlk}. Hence, although simulations with large $a_*$ are strongly influenced by octupole-order effects, the long secular descent timescale ensures that they do not have time to realize the implications of these effects within the age of the universe. It should, in principle, be possible to identify exactly where these regions of hot Jupiter formation lie with detailed phase-space analysis of the ZLK Hamiltonian \citep[e.g.][]{li14chaos}. We defer such an analysis to future work.





\section{Discussion}
\label{section:discussion}

Throughout this analysis, we have examined the formation of double hot Jupiter systems through ZLK migration using a set of fiducial models. Several additional factors must be considered when examining implications for the known population of hot-Jupiter-hosting systems. We consider these factors throughout this section. 

\subsection{Jovian occurrence rates}
In our simulations, we considered binary systems in which a cold Jupiter forms around each star. We have therefore made the assumption that if conditions are amenable to forming a primordial cold Jupiter around one star in a binary pair, the same is true for its companion. 

The observed occurrence rate of cold Jupiters around main-sequence FGK stars is only ${\sim}10-15\%$ \citep{cumming2008keck, wittenmyer2020cool, fulton2021california, gan_2024}, which may, at face value, appear to be in tension with this assumption. However, hot Jupiters are known to preferentially form around metal-rich stars \citep{fischer2005planet, Osborn2019metallicity}, which also host long-period giant exoplanets at a significantly higher rate than the underlying population \citep[e.g.][]{wang2015revealing, ghezzi2018Occurrence, vardan2019giantmetallicity, zink2023hot}. For example, \cite{ghezzi2018Occurrence} shows that, for retired A stars and FGKM dwarfs, the occurrence rate of giant planets grows linearly with respect to the total metal content in the protoplanetary disk, and thus exponentially with metallicity. The two stars within binary systems generally share similar compositional properties because they form from a common molecular cloud \citep{reipurth2007binaries, kratter2011binaries,  king2012multiplicity, vogt2012binaries, Reipurth2012binaryformation, saffee2024chemicaldifferences}. Hence, secondary stars within a population biased toward high metallicities would also likely be high-metallicity and are therefore subject to known correlations between metallicity and giant planet occurrence. 



The occurrence rate of exoplanets around the secondary stars of stellar binary systems is poorly constrained. Only 9\% of known exoplanets in stellar multiples do not orbit the primary, and only in 4 stellar binaries do both components host known circumstellar exoplanets \citep{Kai-Uwe2024gaiacompanion}. However, this absence of detections can be attributed to a combination of biases: for example, transit signals are diluted when both stars fall within a single pixel in large-scale surveys, such as those conducted by the \textit{Kepler} and \textit{TESS} space missions. Secondary stars are also typically smaller and therefore fainter than the primary stars, such that the photometric and radial velocity signal-to-noise ratios are systematically lower. Therefore, while few binary systems are known with an exoplanet detected around \textit{both} stars, this does not necessarily indicate a rarity of such systems. 

The mass of the secondary star may also play an important role in the likelihood that a hot Jupiter may be produced. While the occurrence rate of hot Jupiters is ${\sim}1\%$ around FGK stars \citep{howard2010california, wright2012frequency, beleznay2022exploring, Miyazaki2023sunlike}, hot Jupiters are less common around low-mass M stars \citep{kanodia2024search} and high-mass A stars \citep{beleznay2022exploring}. Observational evidence shows that, more generally, Jovian planets are rare around low-mass M-stars \citep{pass2023mdwarfs}, which would preclude their migration inward to become hot Jupiters. Therefore, in addition to the timescale shifts with stellar mass examined in Section \ref{subsubsection:star-asymmetries}, the underlying occurrence rate of Jovian planets may limit the rate of double hot Jupiter production, particularly for systems in which one or both stars are not FGK type.

\subsection{Binary star system configurations}

The feasibility of ZLK migration is sensitive to the orbital configuration of the binary system in which the hot Jupiters reside. For ZLK oscillations to proceed, a system must include a relatively large mutual inclination $I\gtrsim39.2^{\circ}$ between the inner and outer orbits (e.g. \citet{naoz2016eccentric}; though note that ZLK may be initiated at slightly lower inclinations for the case of a binary perturber \citep{pejcha_2013}). This condition is accounted for within our work, where we examine double hot Jupiter formation under the assumption that orbits are distributed isotropically. 

However, several recent studies have demonstrated that the population of mature planetary systems with close- to moderate- separation ($a\lesssim700$ AU) stellar companions deviates from isotropy, with an excess of low mutual inclinations between exoplanet and stellar binary orbits \citep{christian2022possible, dupuy2022orbital, lester2023visual} that could, in many cases, preclude ZLK evolution. Any trend toward coplanarity post-protoplanetary-disk-dispersal would reduce the rate of expected ZLK-migrated hot Jupiters. Interestingly, the observed trend toward alignment is strongest for non-hot-Jupiter hosts \citep{christian2024wide}, whereas the distribution of hot Jupiter hosts is more suggestive of a mix of coplanarity and ZLK evolution \citep{rice2024orbital}. Therefore, while this alignment trend may affect some subset of the hot-Jupiter-hosting population, it is likely not dominant across the full population, justifying our rough approximation of isotropy.




The physical separation between stars in a binary pair also impacts the likelihood and timescale of double hot Jupiter formation. Demographic surveys have demonstrated that close-separation stellar binaries ($a_*<200$ AU), which undergo relatively rapid ZLK cycles, host $s$-type exoplanets at a lower rate than is found for wide-separation binaries or single-star systems \citep{moe2021impact, hirsch2021understanding}. As a result, such systems may not contain the requisite primordial cold Jupiters needed to produce hot Jupiters through ZLK migration. At the wide-separation end ($a_*> 2000$ AU), ZLK cycles typically proceed too slowly for efficient hot Jupiter formation.


High-eccentricity stellar orbits greatly enhance the efficiency of ZLK migration, and therefore the likelihood of double hot Jupiter formation through the examined mechanism. At the moderate to wide orbital separations relevant to this work, eccentricities of binary star systems are well-described by a flat distribution \citep{duchene2013stellar}. An intriguing population of equal-mass ``twin'' binaries at separations of $a_* >1000$ AU was recently unveiled using astrometric data from the \textit{Gaia} mission \citep{moe2017mind, badry2019discovery}, with a distinct excess of high-eccentricity systems \citep{hwang2022wide}. This ``twin'' binary population is dominated by FGK stars, with a preference for F stars \citep{Obbie2009twinFGK, bakis2020asastwins, yucel2022keplertwins}. At face value, such systems are an excellent match to the types of systems that are most likely to produce ZLK-migrated double hot Jupiters within a short timescale. However, their potentially small pericenter distances at birth -- hypothesized to explain both their near-equal masses and their high eccentricities \citep{tokovinin2020formation, hwang2022wide} -- may lead to suppressed planet formation \citep{jangcondell2008disk} and/or prevent long-term system stability \citep{holman1999long}.

\subsection{Tides}
The efficiency of tidal dissipation is a key parameter dictating the timescales of our model. We emphasize that there are significant uncertainties in planetary tidal theories, both in prescription and precise parameters. We use the simple equilibrium tide theory prescription of \cite{eggleton1998tidespin}, in which hydrostatic equilibrium is assumed and the tidal bulge lags the line of centers by a constant time. In the high-eccentricity epochs of ZLK cycles where the planet undergoes very close pericenter approaches, equilibrium tides may not be the most accurate approach. In particular, dynamical tides consider the excitation of internal modes in response to a tidal perturber, and more accurately account for non-equilibrium dynamics at pericenter (see \cite{mardling_1995, lai_dynamical, ogilvie2014tidal, wu_diffusive, vick_2019}). Dynamical tides predict significantly more dissipation at pericenter, so dynamical tide prescriptions tend to form hot Jupiters significantly faster \citep{vick_2019}. However, as no self-consistent orbital and tidal model exists for dynamical tides at the moment, we defer such an analysis to future work.

Even within the equilibrium tide model, the exact values of relevant tidal parameters are not well known. Constraints on the tidal quality factor $Q$ and tidal Love number $k_2$ of solar system bodies can be inferred from the orbital evolution of their satellites. Through precise astrometric measurements, \cite{lainey2009} constrained $k_2/Q = (1.102 \pm 0.203) \times 10^{-5}$ for Jupiter with data from the Juno mission \citep{bolton2017juno}. Similarly, \cite{lainey2017} found $k_2/Q = (1.59 \pm 0.74) \times 10^{-4}$ for Saturn using Cassini data \citep{spilker2019cassini}. Measurements this precise are not available for exoplanets, and values of the tidal parameters must be indirectly inferred. Exoplanet tidal $Q$ values have been constrained to within an order of magnitude at best \citep[e.g.][]{morley_2017, puranam_chaotic}. This uncertainty in both tidal model and prescription propagates into uncertainties regarding the absolute hot Jupiter formation timescales in our work. However, the relative formation times are unaffected, so our conclusions regarding which systems are most likely to form double hot Jupiters still stand.

\subsection{Ages}
Stellar ages are notoriously difficult to constrain at scale with high precision \citep{tayar2022guide}. The traditional isochrone fitting method has seen great success in constraining ages of star clusters at the population level, but for individual stars -- particularly FGK type dwarfs -- isochrones are so closely separated that precise age constraints are difficult to obtain \citep{pont_isochrone}. Gyrochronology offers more precise constraints through inferences of the star's age from its spin period \citep{barnes_rotational}. Even so, combining gyrochronology and isochrone fitting results in approximately $10\%$ uncertainties at the population level, and higher for individual stars \citep{angus_ages_2019}. These uncertainties in system age naturally propagate into the probability of hot Jupiter formation over some timescale for specific systems.

Throughout this analysis, we halted our simulations when a planet crossed the $q < 0.1$ AU threshold -- in essence, assuming that once a planet becomes a hot Jupiter it remains a hot Jupiter in a stable configuration. This, too, may not necessarily be the case, as some of these planets may be destroyed via tidal disruption or collision with the host star. \cite{levrard_2009} analyzed the orbital configuration of a number of hot Jupiter systems and showed that the vast majority did not have a tidal equilibrium state - implying that further inspiral and future engulfment with the host star was inevitable. \cite{hamer_2019} found that the age of hot Jupiter hosts is on average lower than the field population, further implying that hot Jupiters inspiral and do not survive the main sequence. Observations of individual systems, such as WASP-12 b \citep{yee2020orbit} and Kepler-1658 b \citep{vissapragada_2022_possible}, have supported this claim. Our results do not account for the destruction of inspiraling hot Jupiters, but this would likely decrease the success rate in the highest mutual inclination bins where the hot Jupiters are deposited on tight orbits fairly early in their evolution and have time to decay into the star.

\subsection{Companions to hot Jupiters}

Throughout this work, we have considered systems with only one giant planet around each star. However, previous results have suggested that the observed census of hot Jupiters may be paired with a high rate of massive, wide-orbiting planetary companions \citep{zink2023hot}.

Outer companions contribute a web of competing mechanisms that each impact hot Jupiter formation. On one hand, they may enhance hot Jupiter formation by inducing epochs of planet-planet scattering early in the systems' lifetimes \citep[e.g.][]{rasio1996dynamical, lin1997on, weidenschilling2008gravitational, chatterjee2008dynamical, beauge2012multiple}, resulting in mutual inclinations that potentially make the systems more amenable to either planet-planet or stellar ZLK \citep[e.g.][]{nagasawa2008formation, weldon2025cold}. On the other hand, secular interactions between planets can generate additional precession in the hot Jupiter's orbit, which may produce instabilities in some regions of parameter space \citep{naoz2011hot, denham2019hidden, wei2021relativistic} or which may quench the ZLK effect altogether \citep[e.g.][]{holman1997chaotic,fabrycky2007shrinking}.

Because the orbital properties of the putative wide companions to hot Jupiters are not well-constrained, their impact on the expected double hot Jupiter formation rate remains unclear. We defer such an analysis to future work, acknowledging that, depending on the system properties, the double hot Jupiter formation rate may be either reduced or enhanced by perturbations from such companions. 


\section{Conclusions}
\label{section:conclusions}

In this work, we have quantitatively examined prospects for double hot Jupiter formation -- with one hot Jupiter around each star of a binary system -- through simultaneous ZLK migration. Building from the fiducial case of twin stellar binaries hosting an initial cold Jupiter around each star, we examined the impact of binary separation, asymmetric masses, and varying geometric configuration on the formation of double hot Jupiter systems. Our main conclusions are summarized below.

\begin{itemize}
    \item Using \texttt{REBOUND} \textit{N}-body simulations that account for equilibrium tides and general relativity, we showed that double hot Jupiters are naturally formed for close- to moderate-separation binary systems with sufficiently high mutual orbital inclinations (above the angle required to initiate ZLK oscillations). For close-in, eccentric binary systems $(a_* = 200 \text{ AU}, e_* = 0.7)$, double hot Jupiters formed in nearly $50\%$ of isotropically distributed orbital configurations. In a population synthesis study considering wider-separated binary systems up to $a_*\sim2000$ AU, with an orbital separation distribution motivated by known hot Jupiters with \textit{Gaia}-resolved companions, we found a $9\%$ success rate in forming double hot Jupiters. 

    \item The most favorable binary systems for ZLK-migrated double hot Jupiter formation are those in which the stars have identical, high masses and a relatively close pericenter approach (a few hundred AU) to enable efficient hot Jupiter formation. We recommend, therefore, that a blind search for double hot Jupiters in systems with no previously detected planets should favor binaries with these properties. An overdensity of such binaries has been recently identified using \textit{Gaia} astrometry \citep{moe2017mind, badry2019discovery}; however, follow-up work has suggested that many such systems may be part of a distinct population in which the stars formed in close proximity before being ejected to wider, extremely high-eccentricity orbits \citep{hwang2022wide}. Thus, while these systems constitute an intriguing population for further investigation, their evolutionary pathways may be unfavorable for the initial formation of gas giant planets.

    \item In unequal-mass binary systems, ZLK migration is more efficient around the lower-mass star. Therefore, if a hot Jupiter has formed via ZLK migration around the higher-mass star in a binary pair, and if the secondary star also formed with conditions favorable for the production of ZLK-migrated hot Jupiters (with an initial cold Jupiter, and a sufficiently high mutual inclination), then the second hot Jupiter is likely to have \textit{already} formed around the secondary star. As a result, the lower-mass stars in binary systems with a known hot Jupiter around the higher-mass star are attractive targets for double hot Jupiter search efforts. These include systems such as HAT-P-39 \citep{hartman2012hat,ngo2016friends}, HAT-P-27 \citep{beky2011hat,ngo2016friends} and HAT-P-3 \citep{torres2007hat,stassun2019revised}.
\end{itemize}
Time-series astrometry from \textit{Gaia} DR4 will further demonstrate the feasibility and therefore potential prevalence of hot Jupiter formation through binary-induced ZLK migration. This will arise from two key constraints: an improved characterization of orbital eccentricities for stellar binaries, as well as the discovery of wide-orbiting Jupiter analogs. Both constraints will verify the extent to which the underlying requirements for our model are met -- whether giant planets indeed form around both stars in such binary systems, as well as whether hot-Jupiter-hosting binary star systems are on sufficiently high-eccentricity orbits for efficient ZLK migration.

\section{Acknowledgements}
\label{section:acknowledgements}

We thank the Rice Research Group for helpful conversations over the course of this work. We thank Yubo Su for an insightful discussion. Y.L. acknowledges support from Yale College First-Year Summer Research Fellowship in the Sciences \& Engineering. M.R. and T.L. acknowledge support from Heising-Simons Foundation Grant \#2021-2802. M.R. acknowledges support from Heising-Simons Foundation Grant \#2023-4478 and Oracle for Research grant No. CPQ-3033929. 

This work has benefited from use of the \textit{Grace} computing cluster at the Yale Center for Research Computing (YCRC). This research has made use of the NASA Exoplanet Archive, which is operated by the California Institute of Technology, under contract with the National Aeronautics and Space Administration under the Exoplanet Exploration Program. This work has made use of data from the European Space Agency (ESA) mission
{\it Gaia} (\url{https://www.cosmos.esa.int/gaia}), processed by the {\it Gaia}
Data Processing and Analysis Consortium (DPAC,
\url{https://www.cosmos.esa.int/web/gaia/dpac/consortium}). Funding for the DPAC
has been provided by national institutions, in particular the institutions
participating in the {\it Gaia} Multilateral Agreement.

\software{\texttt{REBOUND} \citep{rein2012rebound}, \texttt{REBOUNDx} \citep{tamayo2019reboundx}, \texttt{matplotlib} \citep{hunter2007matplotlib}, \texttt{numpy} \citep{oliphant2006guide, walt2011numpy, harris2020array}, \texttt{pandas} \citep{mckinney2010data}, \texttt{scipy} \citep{virtanen2020scipy}}



\appendix 
\section{Effect of Planet-Planet Perturbations}
\label{appendix:appendix_perturbations}
Direct \textit{N}-body simulations capture all relevant dynamics of the systems examined in this work, but they are computationally expensive for population-level ensembles. Orbit-averaged secular codes are significantly faster and hence more computationally tractable; however, precise models of the three-body ZLK effect may not well describe the behavior of binary systems with two planets, particularly in the regime of close approaches between the two stellar systems. In this section, we quantitatively assess the degree to which the standard three-body ZLK effect can approximate the system dynamics at the population level. We identify the regime in which this approximation fails to capture the system dynamics and justify the use of four-body \textit{N}-body simulations for our purposes.

We again use \texttt{REBOUND} to numerically quantify the perturbative influence of the neighboring planet. We run and compare two ensembles: a set of four-body simulations with a planet initialized around each star, and a set of three-body simulations with a planet only around one star. We adopt the fiducial parameters from Table \ref{tab:fiducial}, but set the binary separation to $a_*=500$ AU, $a_*=600$ AU, and $a_*=700 \mathrm{AU}$ and eccentricity to $e_*=0.1$, $e_*=0.5$, $e_*=0.9$ along a grid to simulate more ``typical'' systems. We vary the inclination of the planet of interest from $75^\circ$ to $105^{\circ}$, and we keep the neighboring planet's inclination fixed at $97^\circ$. In the three-body simulation, all system parameters are exactly the same as in the four-body setup, but rather than including the planet with $97^\circ$ inclination, the corresponding host star is endowed with mass $1 {M_\odot} + 1 {M_{\rm Jup}}$ to keep the total perturbing mass constant. 

\begin{figure}
    \centering
    \includegraphics[width=0.48\textwidth]{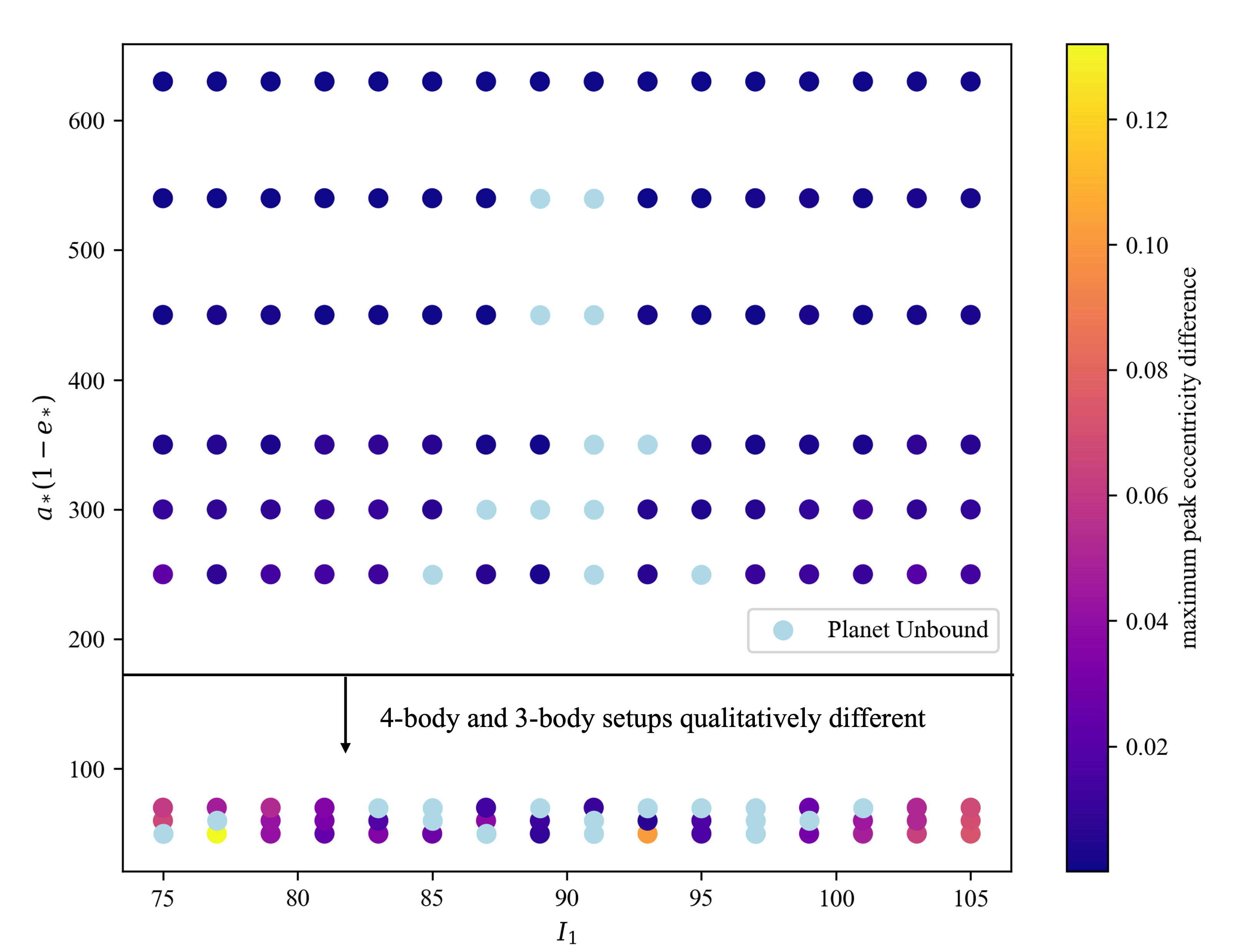}
    \caption{Maximum planetary orbit peak eccentricity difference between (1) three-body binary star systems with one planet and (2) four-body binary star systems with two planets. Light blue points indicate that at least one simulation in the pair produced an unbound planet. The parameter space with binary pericenter $a_*(1-e_*)<200$ AU or initial planetary inclination $85^{\circ}<I_1<95^{\circ}$ exhibits significant deviations in peak eccentricities due to perturbations from the companion planet.}
    \label{fig:perturbation}
\end{figure}

Our simulations show that some systems, shown in light blue in Figure \ref{fig:perturbation}, become unstable, resulting in an unbound planetary orbit. The three-body system is a poor approximation for the four-body system in this case.

For stable systems, we compare the eccentricity evolution of the planet shared by the four-body and three-body setups for each pair of simulations. For each pair of corresponding peaks—i.e., the $i^{th}$ peak in both simulations—we calculate the difference in peak eccentricity. We use the maximum peak eccentricity difference $\Delta e_{\rm peak}$ to quantify the extent to which the two simulations differ, because the peak eccentricity drives the rate of tidal dissipation. As can be seen in Figure \ref{fig:perturbation}, the parameter space is divided into two regions. When the stellar pericenter is less than 200 AU or when $85^\circ<I_1<95^\circ$, $\Delta e_{\rm peak}$ is noticeably larger or systems regularly become unstable. The maximum peak eccentricity difference in our simulations reaches values as high as $\Delta e_{\rm peak}=0.13$. In this region, we observe clear qualitative differences in the shape and amplitude of the eccentricity oscillations of the four-body simulation as compared with the three-body simulation. In such chaos-driven cases, the three-body simulation fails to well-approximate the behavior of the full system. Notably, this region—characterized by high initial inclination and low stellar pericenter—is where ZLK-driven hot Jupiter formation is most favored. Consequently, orbit-averaged three-body simulations are unsuitable for our purposes.

\begin{figure}
    \centering
    \includegraphics[width=0.48\textwidth]{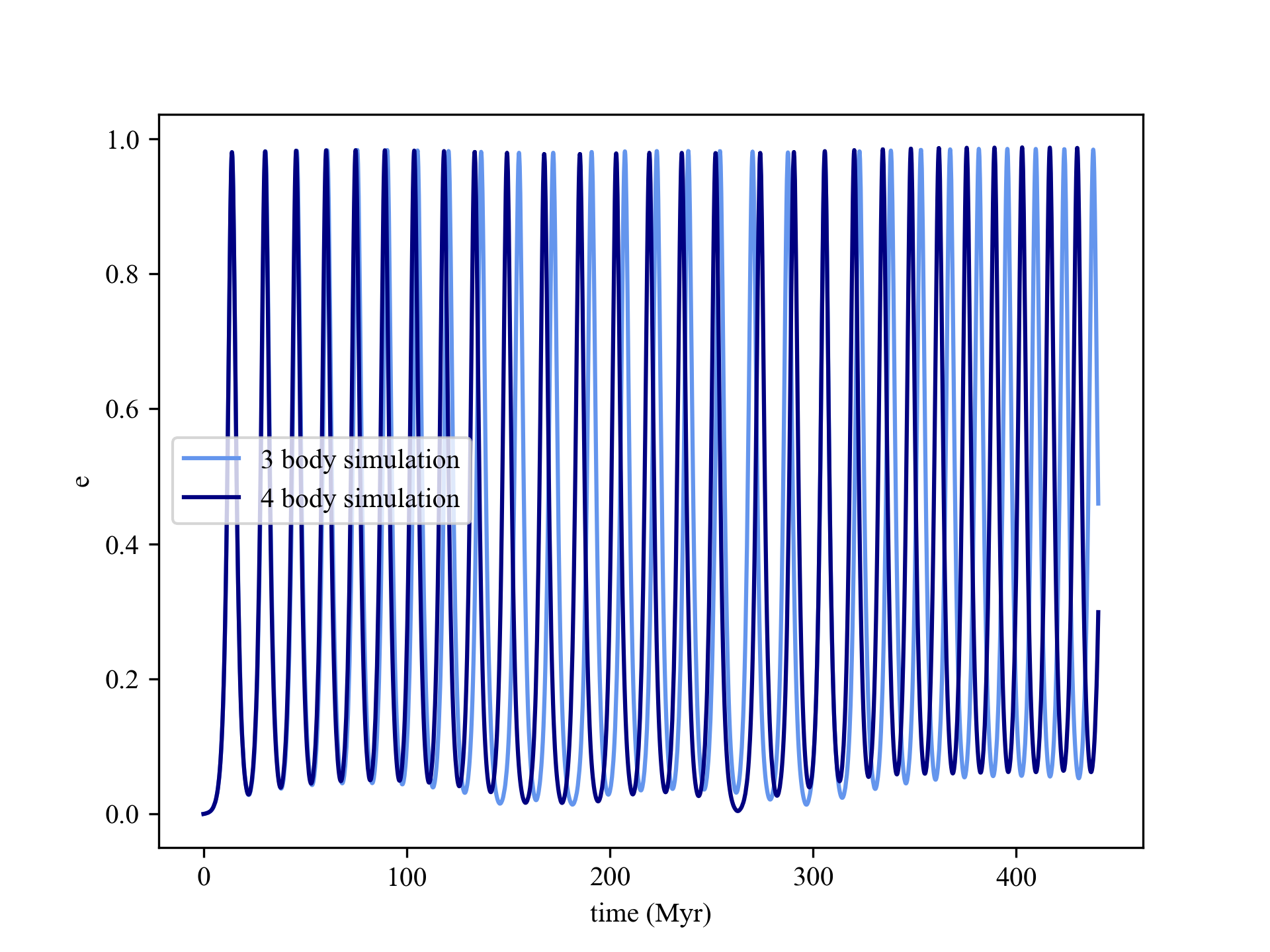}
    \caption{Comparison of eccentricity evolution between the traditional three-body ZLK setup, with only one planet, and our fiducial 4-body simulation with two planets. System parameters are set to $a_*=700 \mathrm{AU}$, $e_*=0.5$, $I_1 = 99^{\circ}$. The companion planet produces a phase shift in the ZLK oscillations, leading to a timing offset but no significant amplitude difference.}
    \label{fig:phase_shift}
\end{figure}

For systems with pericenter distance larger than 200 AU and $I_1<85^\circ$ or $I_1>95^{\circ}$, the difference in eccentricity evolution is less pronounced. For these systems, the period of ZLK oscillations differs between the four-body and three-body simulations, leading to a phase offset but no qualitative differences as in the chaotic region. Figure \ref{fig:phase_shift} shows an example of a system that behaves as such. In this region of parameter space, the peak eccentricities of the two simulations differ only slightly: across our suite of simulations, we find a maximum peak eccentricity difference $\Delta e_{\rm peak}=0.023$ in this regime. Therefore, we consider these systems to be well-approximated by the equivalent three-body system evolution. We note, however, that this peak eccentricity difference, though small, affects the energy dissipation rate of the planet at the closest approach to an order as high as $e^2$ as given in Equation (36) in \cite{millholland2019obliquity}. Thus, while systems in this parameter space are relatively ``well-behaved'' by comparison with their closer-separation counterparts, the three- and four-body cases may still demonstrate some qualitative differences.

A phase offset may also interfere with the appropriateness of using a three-body system to approximate a four-body system. If double hot Jupiter systems take considerably longer to form than single hot Jupiter systems, then some four-body systems might not be able to produce hot Jupiters within the system's lifetime, while the corresponding three-body systems could. Conversely, in some simulations, the three-body system takes longer to create a hot Jupiter, and a system may mistakenly be marked as a ``failed'' case when it could produce double hot Jupiters. To quantify the effect of phase offsets, we calculate the ratio $T_4/T_3$ between the time the four-body vs. three-body system takes to reach a given number of eccentricity peaks. In the less chaotic region of Figure \ref{fig:perturbation}, the maximum deviation $|T_4/T_3 - 1|$ is 0.19, indicating that, in the parameter ranges examined within this work, the formation time of hot Jupiters can differ by up to 19\% by approximating four-body system evolution with three bodies.

\bibliography{bibliography}
\bibliographystyle{aasjournal}

\end{CJK*}
\end{document}